\documentclass{appolb}
\usepackage{graphicx}

\usepackage{amssymb}

\begin{document}
\title{Theory input for $t\bar{t}j$ experimental analyses at the LHC%
\thanks{Presented at Matter To The Deepest (MTTD) 2021}%
}
\author{Katharina Vo{\ss}$^{\mathrm{a},\mathrm{b}}$, Maria Vittoria Garzelli$^{\mathrm{a}}$, Sven-Olaf Moch$^{\mathrm{a}}$
\address{$^\mathrm{a}\,$II. Institut f\"ur Theoretische Physik, Universit\"at Hamburg, \\ Luruper Chaussee 149, D~--~22761 Hamburg, Germany\\
$^{\mathrm{b}}\,$Department f\"ur Physik,  Universit\"at Siegen, 
  Emmy Noether Campus, \\ Walter Flex Str. 3, D~--~57068 Siegen, Germany}}

{\small December 2021 \hfill DESY 21-229} \\
{\let\newpage\relax\maketitle}
\begin{abstract}
The precise measurement of the top quark mass, which is a fundamental SM parameter, constitutes one of the main goals of the LHC top physics program. One approach to measure this quantity uses the $\rho_\mathrm{s}$ distribution, an observable depending on the invariant mass of the $t\bar{t}j$ system. To fully exploit the experimental accuracy achievable in measuring top quark production cross sections at the LHC, the theory uncertainties associated to these measurements need to be well under control. To this end we present a study of the effect of varying the theoretical input parameters in the calculation of differential cross sections of the $t\bar{t}j$ process. Thereby we studied the influence of the jet reconstruction procedure, as well as the effect of various renormalization and factorization scale definitions and different PDF sets.  The variation of the $R$ parameter in the jet reconstruction algorithm was found to have negligible influence on the scale variation uncertainty.  A strong reduction of scale uncertainties and a better behaviour of the NLO/LO ratios using selected dynamical
scales instead of a static one in the high energy tails of differential distributions was observed. This is particularly interesting in the context of the top quark mass measurements through the $\rho_\mathrm{s}$ distribution, in which the perturbative stability can be improved by applying the proposed dynamical scale definition.
\end{abstract}
  
\section{Introduction}
\label{sec:intro}
The precise measurement of the properties and interactions of the top-quark, the fundamental particle with the largest mass in the Standard Model (SM), is possible due to the high statistics in top quark production processes available at the Large Hadron Collider (LHC). Besides enabling the investigation of the top-quark Yukawa coupling, the largest Yukawa-coupling in the SM, and testing Beyond the Standard Model theories, measurements of the top-quark SM parameters with unprecedented accuracy can be carried out. In particular, the top-quark mass is linked to fundamental questions in particle physics. For example, the mass of the top-quark $m_t$, the Higgs-Boson $m_H$ and the $W$-Boson $m_W$ are linked through radiative corrections and these mutual dependencies can be exploited to perform a consistency check of the predictions of the SM~\cite{Awramik:2004}. Furthermore, the stability or meta-stability of the electroweak vacuum can be inferred from the relations between $m_t$, $m_H$ and the strong coupling constant $\alpha_s$ \cite{Bezrukov:2012,Degrassi:2012,Alekhin:2012,Bednyakov:2015}.\\
The possibility to extract the top-quark mass through the normalized $\rho_\mathrm{s}$-distribution, defined as $\rho_\mathrm{s} = 2 m_0 / \sqrt{m_{t\bar{t}j}^2}$ with $m_0=170\,$GeV, was first discussed in \cite{Alioli:2013}. This quantity, calculated at NLO in the $t\bar{t}j$-process, was shown to be more sensitive to $m_t$ than the corresponding distribution of $2 m_0 / \sqrt{m_{t\bar{t}}}$ for the $t\bar{t}$-process. In fact, additional parton radiation in the $t\bar{t}j$-process gives enhanced sensitivity to the mass of the top-quark. As in the extractions of the top-quark mass from cross section measurements, the mass renormalization scheme can be unambiguously defined, when measuring $m_t$ through the $\rho_\mathrm{s}$ distribution. Thereby, in \cite{Alioli:2013} it was found that the $\rho_\mathrm{s}$ distribution is more sensitive to $m_t$ than the cross section measurements.\\
Both ATLAS~\cite{ATLAS:2019JHEP} and CMS~\cite{CMS-PAS-TOP-13-006} performed measurements of the top-quark mass using the $\rho_\mathrm{s}$ distribution, where the most recent ATLAS result~\cite{ATLAS:2019JHEP} extracted the pole mass of the top-quark with the value
\begin{equation}
m_t^{\mathrm{pole}} = 171.1 \pm 0.4\,\mathrm{(stat)} \pm 0.9\,\mathrm{(syst)} ^{+0.7}_{-0.3}\,\mathrm{(theo)}\,\mathrm{GeV}.
\end{equation}
The theoretical uncertainty in the extraction is sizeable, which is dominated by the scale variation uncertainty $(+0.6,-0.2)\,\mathrm{GeV}$, while the parton distribution function (PDF) and $\alpha_s$ uncertainty leads only to an uncertainty of $\pm 0.2\,$GeV in the $m_t^{\mathrm{pole}}$-determination. Therefore, we carried out an investigation of the theoretical uncertainties in the $t\bar{t}j$-process and the possibility to improve the 
perturbative stability through informed choices of the theoretical input in the calculation and in the analysis procedure. \\
A study of the $t\bar{t}j$ production process is furthermore interesting, since a substantial fraction of $t\bar{t}$-events at the LHC are accompanied by an additional jet (40\,\% of $t\bar{t}$-events, if $p_T^j>40\,$GeV at $\sqrt{s}=13\,$TeV \cite{Kraus:2016}) and this process constitutes the dominant background to Higgs production in vector boson fusion (VBF), see e.g. \cite{Rainwater:1999}.\\
The first NLO calculation of $t\bar{t}j$ production at a hadron collider was presented in \cite{Dittmaier:2009}, where stable top-quarks were considered. In \cite{Melnikov:2010} the LO top-quark decay was included and later in \cite{Bevilacqua:2016} the NLO QCD off-shell effects in the fully leptonic decay of the top-quark were considered. The combination of the NLO $t\bar{t}j$ calculation and a parton shower (PS) was first presented in \cite{Kardos:2011} using hard scattering amplitudes from HELAC-NLO \cite{vanHameren:2009} and the PS matching method implemented in the POWHEG-BOX \cite{Alioli:2010}. A second implementation, using the virtual corrections from \cite{Dittmaier:2009} and Born and real squared amplitudes from Madgraph \cite{Alwall:2007}, in the POWHEG-BOX, called in the following \texttt{ttbarj V1} \cite{Alioli:2012} was published shortly afterwards. In the study presented here, the \texttt{ttbarj V2} version was used, where in contrast to \texttt{ttbarj V1}, all amplitudes are calculated with \texttt{OpenLoops2}~\cite{Buccioni:2019} and the calculation can be parallelized, leading to strongly reduced computation time. 

\section{Input parameters of the calculation}
\label{sec:input}
The presented predictions are obtained with a NLO calculation of the $t\bar{t}j$ process at a center of mass energy of $\sqrt{s}=13\,$TeV. The pole mass of the top-quark is set to $m_t^{\mathrm{pole}}=172\,$GeV and stable top-quarks are considered. As a PDF set and for the evolution of the strong coupling constant $\alpha_s$ the CT18NLO PDF set \cite{Hou:2019} was used as default. To estimate the scale variation uncertainty the seven point scale variation method was used, varying the renormalization $\mu_R = K_R \mu_0$ and factorization scale $\mu_F = K_F \mu_0$ in the range
\begin{equation}
(K_R,K_F) \in \{ (0.5,0.5), (0.5,1), (1,0.5), (1,1), (1,2), (2,1), (2,2) \}.
\end{equation}
In the analysis, at least one jet with a transverse momentum $p_T^j > 30\,$GeV and a pseudorapidity $|\eta_j|<2.4$ was required, where the jets are reconstructed with the anti-$k_T$ jet clustering algorithm from \texttt{FastJet}~\cite{Cacciari:2012} using the $E$-recombination scheme and the $R=0.4$ value. \\
In the calculation we used different definitions of the central scale, as it was found in \cite{Bevilacqua:2016} that dynamical scales were able to better describe the high-energy tails of various NLO differential distributions for $pp \rightarrow t\bar{t}j$ in fully leptonic decay compared to the static scale $\mu_0=m_t^{\mathrm{pole}}$. In contrast to \cite{Bevilacqua:2016}, our study was carried out with more inclusive analysis cuts and stable top-quarks. This is owed to the refined techniques of the experimental collaborations to unfold to parton level. We considered four scale definitions, a static scale $\mu_0=m_t^{\mathrm{pole}}$ and three dynamical scales $\mu_0 = m_{t\bar{t}j}^B/2, H_T^B/2$ and $H_T^B/4$, where
\begin{equation}
m_{t\bar{t}j} = \sqrt{(p_t^B + p_{\bar{t}}^B + p_j^B)^2}, \hspace{0.2cm}
H_T^B = \sqrt{{p_{T,t}^B}^2 + {m_t^{\mathrm{pole}}}^2} + \sqrt{{p_{T,\bar{t}}^B}^2 + {m_t^{\mathrm{pole}}}^2} + p_{T,j}^B.
\end{equation}
The superscript $B$ implies that the kinematic variables are reconstructed at the underlying Born level in the POWHEG-BOX.

\section{Evaluation of the scale variation uncertainty}
\subsection{Scale variation uncertainty in the $\rho_\mathrm{s}$ distribution}
\label{sec:scalevar}
\label{subsec:scalevar_rho}
\begin{figure}[htb]
\centerline{%
\includegraphics[width=12.5cm]{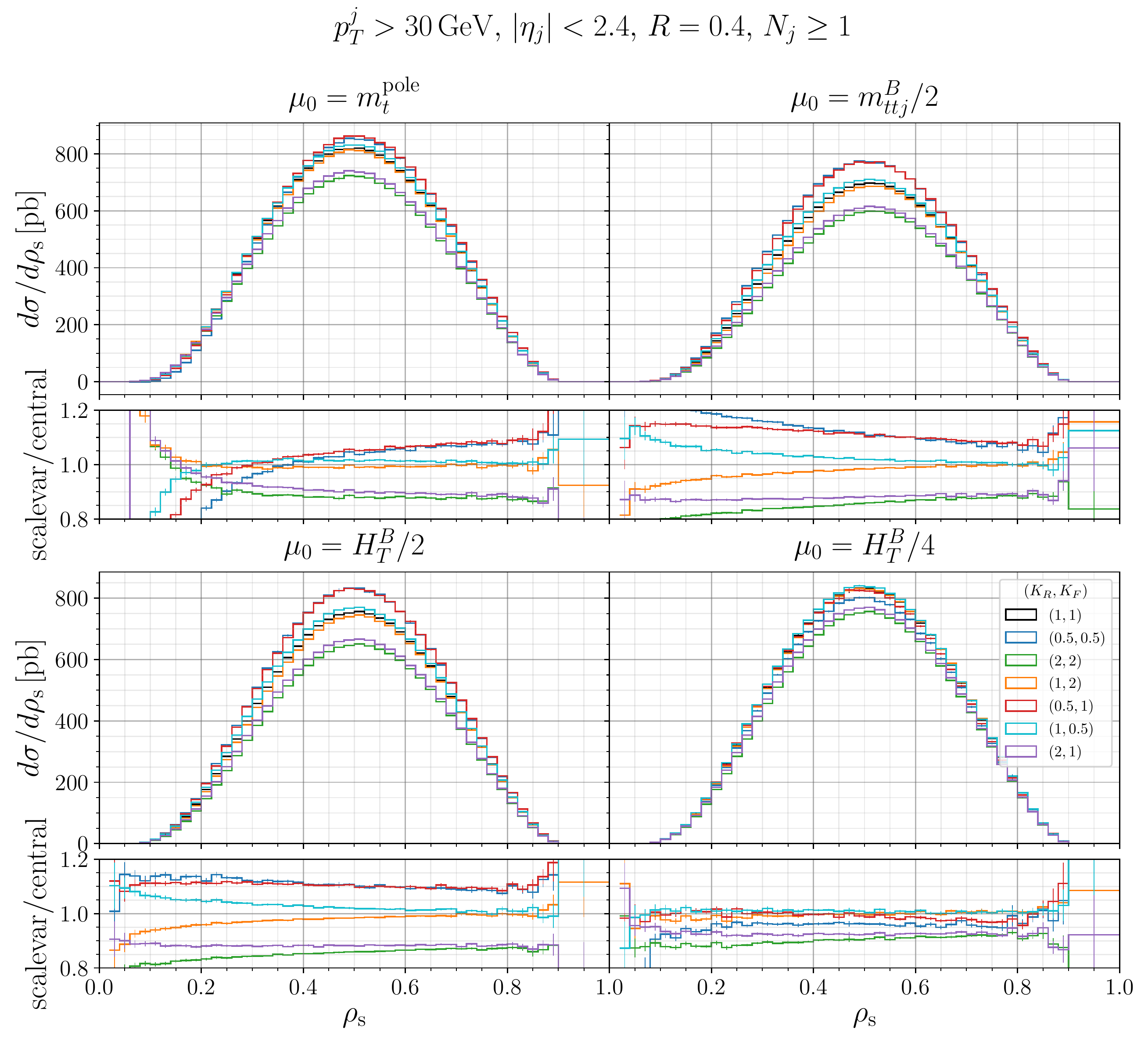}}
\caption{NLO differential cross section of the process $pp \rightarrow t\bar{t}j +X$ at $\sqrt{s}=13\,$TeV as a function of $\rho_\mathrm{s}$ obtained with the scales $\mu_0=m_t^{\mathrm{pole}}$, $m_{t\bar{t}j}^B/2$ (top, from left to right), $H_T^B/2$ and $H_T^B/4$ (bottom, from left to right). The seven scale variation graphs are drawn explicitly, while in each of the lower insets the ratios of these with the prediction obtained with $K_R=K_F=1$ (black) are shown.}
\label{Fig:rho}
\end{figure}

\begin{figure}[htb]
\centerline{%
\includegraphics[width=11cm]{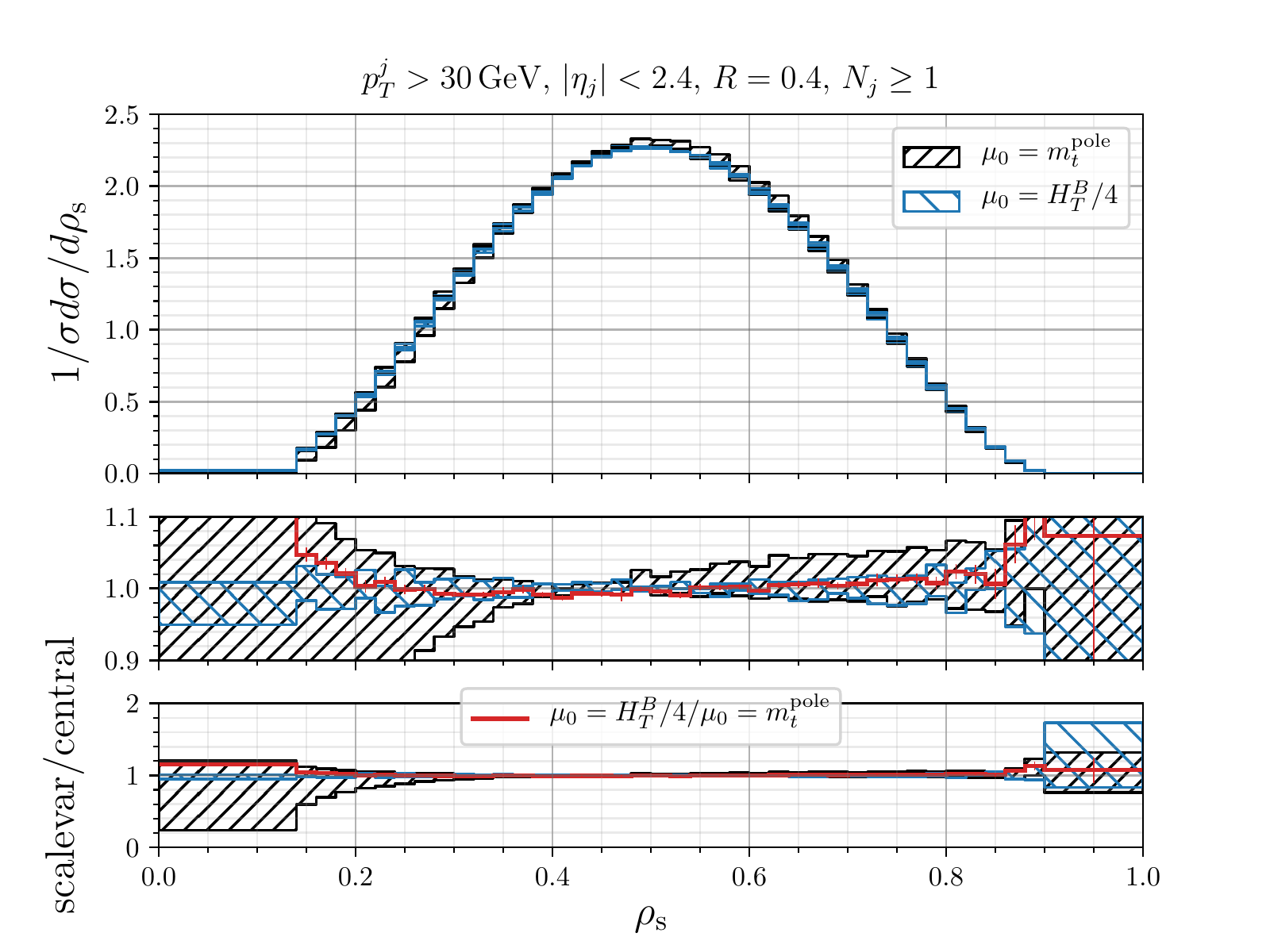}}
\caption{NLO differential cross section of the process $pp \rightarrow t\bar{t}j +X$ at $\sqrt{s}=13\,$TeV and the seven point scale uncertainty band as a function of $\rho_\mathrm{s}$ obtained with the static scale $\mu_0=m_t^{\mathrm{pole}}$ (black) and the dynamical scale $\mu_0=H_T^B/4$ (blue). The ratio plots in the lower two panels show the scale variation uncertainty bands normalized to the distribution obtained with $K_R=K_F=1$, while using the corresponding scale definition. Thereby, the two ratio plots differ only in the displayed $y$-axis range. The ratio of the two nominal predictions is shown by the red graph.}
\label{Fig:rho_mt_ht4}
\end{figure}

\begin{figure}[htb]
\centerline{%
\includegraphics[width=12.5cm]{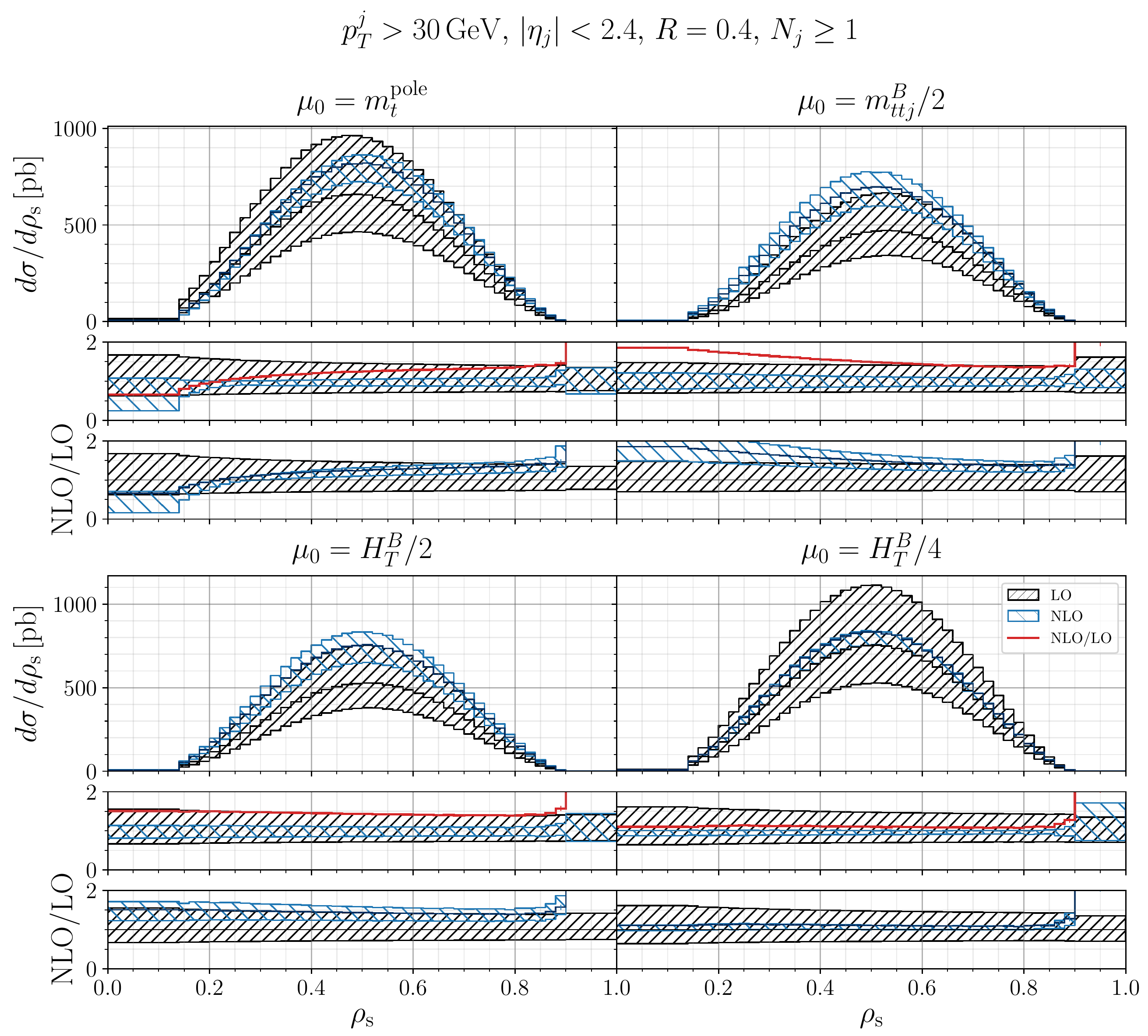}}
\caption{NLO (blue) and LO (black) differential cross section of the process $pp \rightarrow t\bar{t}j +X$ at $\sqrt{s}=13\,$TeV as a function of $\rho_\mathrm{s}$ obtained with the scales $\mu_0=m_t^{\mathrm{pole}}$, $m_{t\bar{t}j}^B/2$ (top, from left to right), $H_T^B/2$ and $H_T^B/4$ (bottom, from left to right). The seven point scale variation uncertainty bands are drawn, which are rescaled in the middle ratio plot to either the LO or NLO central scale prediction, while in the lower ratio plot both NLO and LO scale variation bands are rescaled by the LO central scale prediction.}
\label{Fig:rho_nlo_lo}
\end{figure}

As the $\rho_\mathrm{s}$ distribution provides a possibility to measure the top-quark mass and the theoretical uncertainty in this extraction is sizeable, as anticipated in Section~\ref{sec:intro}, this distribution is discussed in detail in the following. \\
In Fig.~\ref{Fig:rho} the NLO $\rho_\mathrm{s}$ distribution is shown for the static scale $\mu_0=m_t^{\mathrm{pole}}$ and the dynamical scales $\mu_0 = m_{t\bar{t}j}^B/2, H_T^B/2$ and $H_T^B/4$. Besides the prediction obtained with $\mu_R = \mu_F = \mu_0$ also the six scale variation graphs, generated by varying $K_R$ and $K_F$ appearing in $(\mu_R,\mu_F) = (K_R,K_F)\mu_0$ with values $K_R, K_F \in \{ 0.5, 1, 2\}$, leaving out the extreme combinations (0.5,2) and (2,0.5), are shown explicitly. In case of the static scale $\mu_0=m_t^{\mathrm{pole}}$ (left panel) a large spread in the graphs, from which the scale variation uncertainty band is built, is observed at low $\rho_\mathrm{s}$, which corresponds to large values of $m_{t\bar{t}j}$ and as such to the high energy region. Furthermore, a crossing of the scale variation bands occurs using the static scale for $0.1 \lesssim \rho_\mathrm{s} \lesssim 0.3$. Such a pronounced behaviour is not seen using the dynamical scales and the scale variation induces a smaller shape variation of the $\rho_\mathrm{s}$ distribution in these cases.
This leads to a strongly reduced scale variation uncertainty, when comparing the normalized $\rho_\mathrm{s}$ distributions. In this case, each distribution obtained with different $\mu_R, \mu_F$ values is normalized by the total cross section calculated with these input values. As an example, the normalized $\rho_\mathrm{s}$ distribution including the seven point scale variation uncertainty band is shown in Fig.~\ref{Fig:rho_mt_ht4} for the static scale $\mu_0=m_t^{\mathrm{pole}}$ (black) and the dynamical scale $\mu_0=H_T^B/4$ (blue), which was found to be the dynamical scale leading to the smallest scale variation uncertainty. The strong reduction of the scale variation uncertainty bands is especially visible in the lower two ratio plots, in which the scale variation bands are rescaled by the corresponding nominal distribution. Besides showing a smaller scale uncertainty in the region of low $\rho_\mathrm{s}$ values, in which the description with the static scale is problematic, a smaller uncertainty band is obtained also in the bulk of the distribution.\\
A further possibility to evaluate the most preferable choice of the scale, is to compare the calculation of the $\rho_\mathrm{s}$ distribution at different perturbative orders, explicitly in this case the $\rho_\mathrm{s}$ distribution calculated at LO and NLO. A central scale choice, which leads to small high-order corrections is desirable, as it can indicate that the calculation will not deviate from the current NLO result in higher, yet uncalculated orders. This comparison, including also the seven point scale variation uncertainty bands at either LO (black) or NLO (blue), is shown in Fig.~\ref{Fig:rho_nlo_lo}. Especially comparing the lower ratio plots, in which both seven point scale variation bands are rescaled by the LO central scale prediction, the two scales $\mu_0=m_t^{\mathrm{pole}}$ and $\mu_0=m_{t\bar{t}j}^B/2$ seem problematic, since the NLO and LO scale variation bands only slightly overlap in the high-energy region. On the other hand, the description using $\mu_0=H_T^B/2$ and $\mu_0=H_T^B/4$ shows a more uniform differential $\mathcal{K}$-factor, depicted with the red line in the middle plot of Fig.~\ref{Fig:rho_nlo_lo}. The differential $\mathcal{K}$-factor using $\mu_0=H_T^B/4$ is close to one, which was the original motivation to investigate this dynamical scale, since the ratio of the integrated cross section calculated at NLO and LO was also found to be near one. A similar behaviour in the high-energy tails of differential distributions was also found for other observables and agrees with the findings in \cite{Bevilacqua:2016}.
\subsection{Influence of the $R$-parameter on the scale variation uncertainty}
Investigating further the static scale $\mu_0=m_t^{\mathrm{pole}}$ and the two dynamical scales, which seem to be preferable considering the results of Section~\ref{subsec:scalevar_rho}, namely $\mu_0=H_T^B/2$ and $\mu_0=H_T^B/4$, the influence of the choice of the $R$-parameter in the anti-$k_T$ jet clustering algorithm is studied. Two values of this parameter are considered, the default value of $R=0.4$ and an additional value of $R=0.8$, for which the experimental analyses have also investigated reconstruction efficiencies and Monte Carlo to data comparison in order to determine the systematic uncertainties. \\
Using the dynamical scales, the choice of the $R$ parameter has only a minor influence on the size of the scale variation uncertainty, while the predictions obtained with the static scale show a slightly reduced scale uncertainty in the high-energy tails with the larger $R=0.8$ value. However, as elaborated in the previous section, this phase space region seems not to be well described using the static scale. Using either scale definition, the differential cross sections show larger values using the larger $R$ parameter, which can be advantageous in statistically limited analyses. The same features, described here for the $\rho_\mathrm{s}$ distribution, were also found for several other differential cross sections. 
\begin{figure}[htb]
\centerline{%
\includegraphics[width=12.5cm]{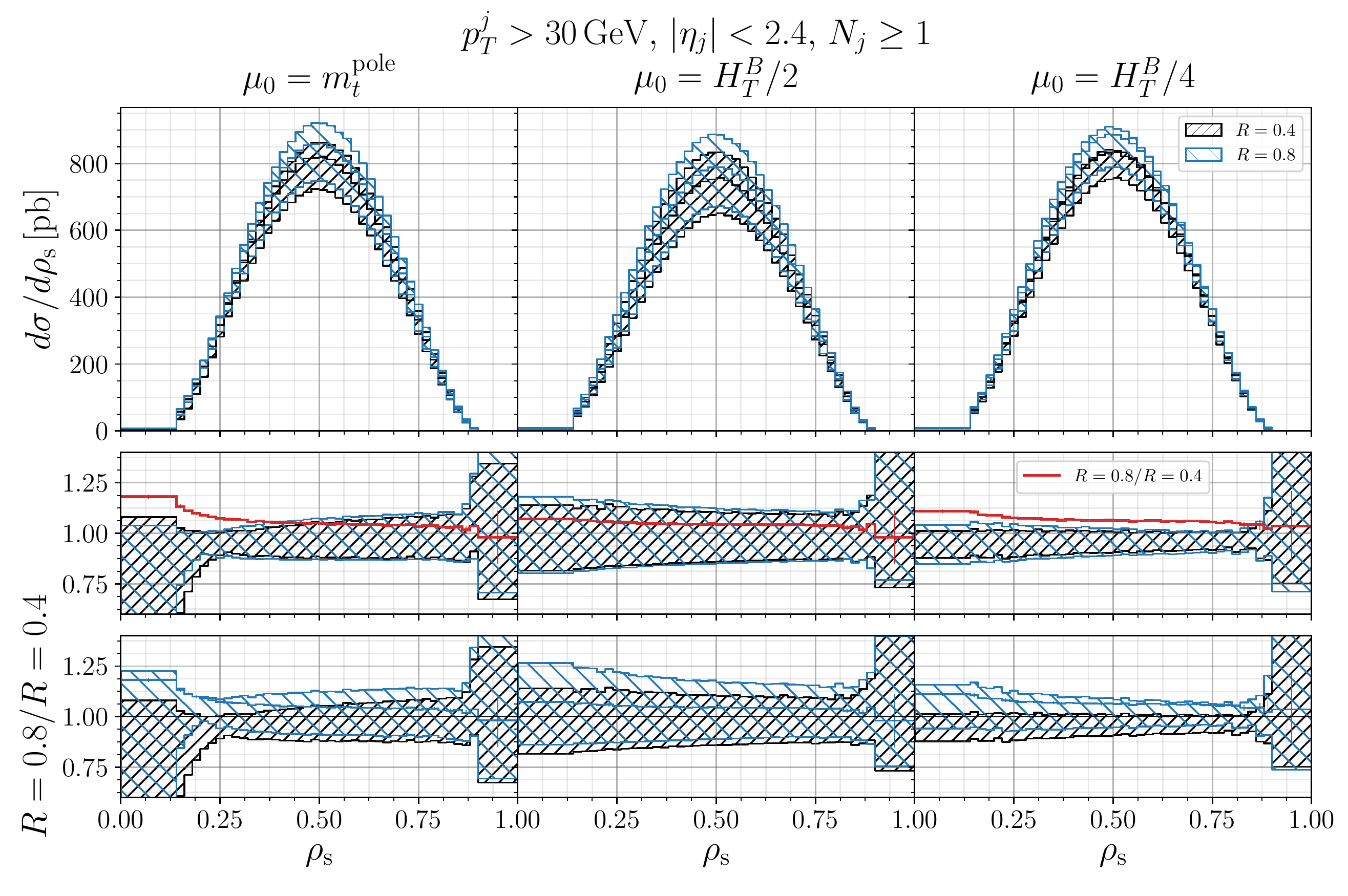}}
\caption{NLO differential cross section of the process $pp \rightarrow t\bar{t}j +X$ at $\sqrt{s}=13\,$TeV as a function of $\rho_\mathrm{s}$ obtained with the scales $\mu_0=m_t^{\mathrm{pole}}$, $H_T^B/2$ and $H_T^B/4$ (from left to right) using the anti-$k_T$ jet clustering algorithm with either $R=0.4$ (black) or $R=0.8$ (blue). The seven point scale variation uncertainty band is drawn, which is rescaled in the middle ratio plot to either central scale prediction (using either $R$ value), while in the lower ratio plot both scale variation bands are rescaled by the central scale prediction obtained by setting $R$ to the default value of $R=0.4$.}
\label{Fig:rho_Rvar}
\end{figure}

\section{Evaluation of the PDF uncertainty}
\label{sec:pdf}
\begin{figure}[htb]
\centerline{%
\includegraphics[width=11cm]{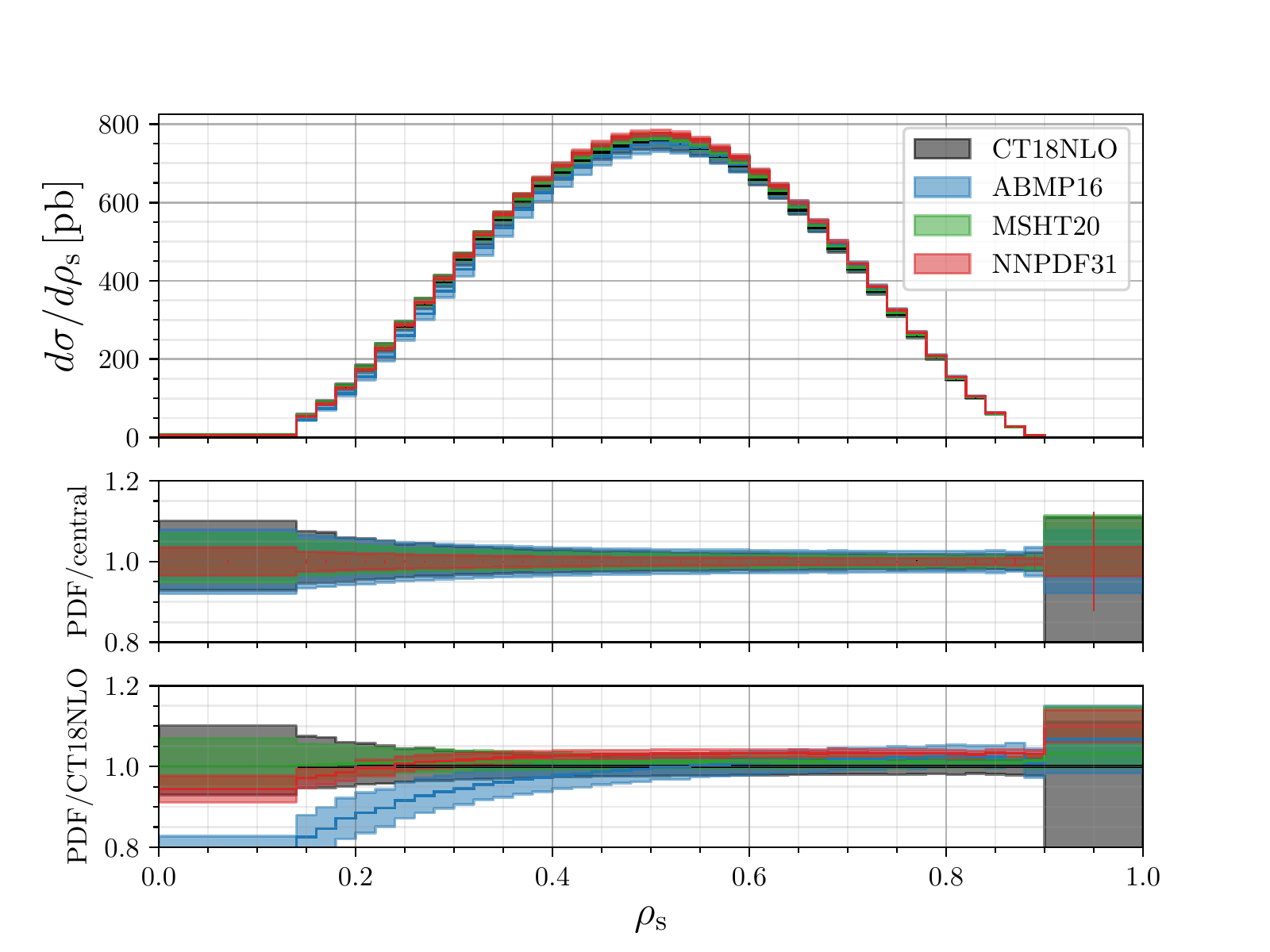}}
\caption{Approximation of the NLO PDF uncertainties, as described in Section~\ref{sec:pdf}, for the $\rho_\mathrm{s}$ distribution of the process $pp \rightarrow t\bar{t}j +X$ at $\sqrt{s}=13\,$TeV obtained with the dynamical scale $\mu_0=H_T^B/4$. Four NLO PDF sets were studied, CT18NLO (black), ABMP16 (blue), MSHT20 (green) and NNPDF3.1 (red) and the PDF uncertainties calculated as recommended by the authors of the corresponding PDF fits.}
\label{Fig:rho_pdfvar}
\end{figure}

\begin{figure}[htb]
\centerline{%
\includegraphics[width=12.5cm]{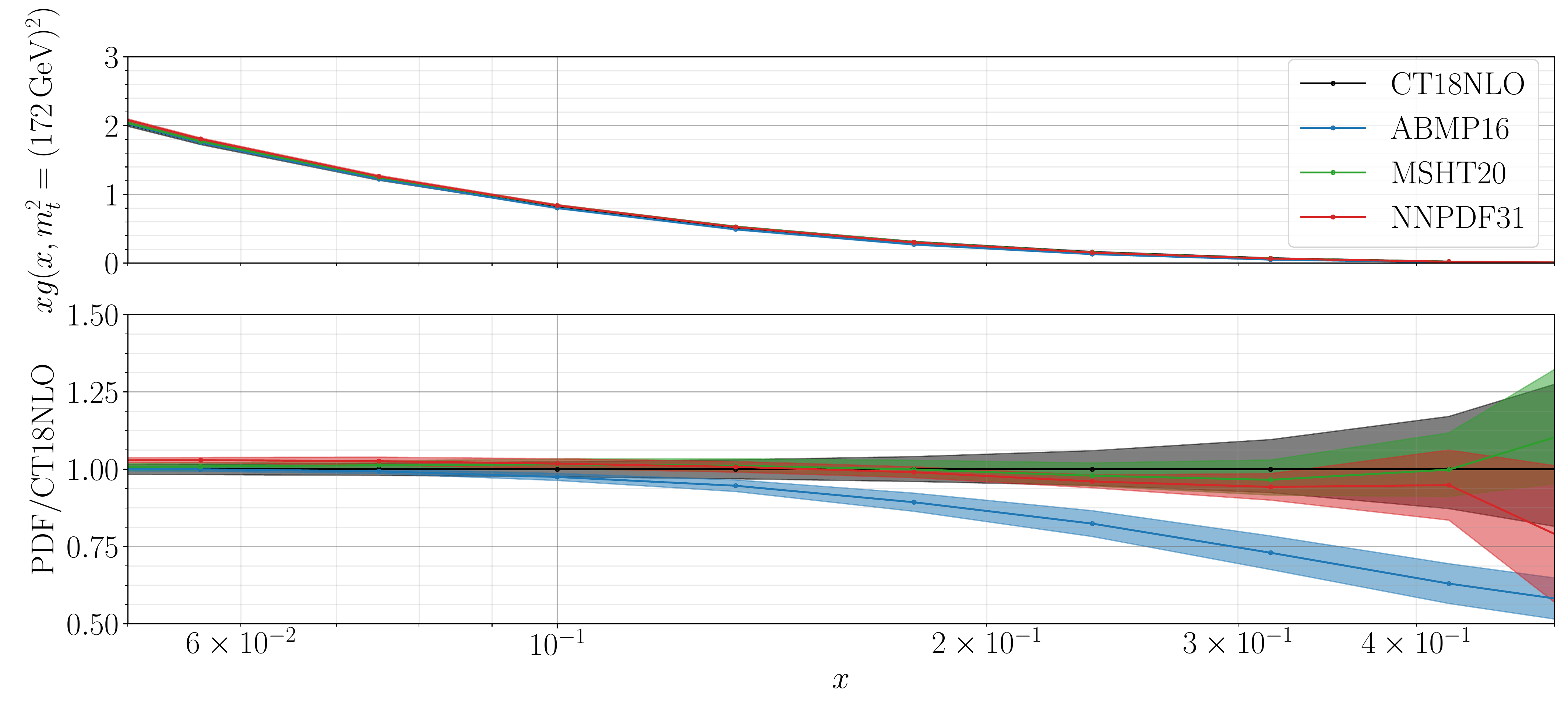}}
\caption{Gluon PDFs as a function of the momentum fraction $x$, while setting $Q^2=({m_t^{\mathrm{pole}}})^2 = (172\,\mathrm{GeV})^2$. Four NLO PDF sets were considered, CT18NLO (black), ABMP16 (blue), MSHT20 (green) and NNPDF3.1 (red) and the PDF uncertainties calculated as recommended by the authors of the corresponding PDF fits.}
\label{Fig:xgluon_pdfvar}
\end{figure}

As the predictions obtained with the scale choice $\mu_0=H_T^B/4$ were found to have desirable features in Section~\ref{sec:scalevar}, leading to a small scale variation uncertainty, overlapping NLO and LO scale variation bands and a stable description of the high-energy region, the PDF uncertainties using this same scale, are investigated in the following. Thereby four modern PDF sets are compared, i.e. the default PDF set used in this study CT18NLO and the ABMP16 \cite{Alekhin:2018}, MSHT20 \cite{Bailey:2021} and NNPDF3.1 \cite{Ball:2017} NLO sets. The PDF uncertainties were calculated according to the recommendations given by the authors of each PDF fit. Due to the large computational effort needed to calculate the corresponding distributions for each eigenvector set of the different PDF fits, the approximation of using LO partonic cross sections in combination with NLO PDF sets instead of NLO partonic cross sections to estimate the NLO PDF uncertainty, is used in the presented predictions. The validity of this approximation was tested, by comparing the percentage size of the PDF uncertainty bands in several differential distributions using the NLO and LO partonic cross section and finding very good agreement. This study was performed using as a central scale $\mu_0=H_T^B/2$ and a further comparison of the percentage size of the PDF uncertainty bands obtained with $\mu_0=H_T^B/4$ and $\mu_0=H_T^B/2$ also showed very similar results. \\
In Fig.~\ref{Fig:rho_pdfvar} we present the $\rho_\mathrm{s}$ distribution calculated with LO partonic matrix elements using $\mu_0=H_T^B/4$ and the four different NLO PDF sets described above to approximate the NLO PDF uncertainties. In the bulk of the $\rho_\mathrm{s}$ distribution good agreement between the predictions from the different PDF sets is observed, while in the region of low $\rho_\mathrm{s}$ differences are seen, which are not covered by the PDF uncertainty bands. This was found to be caused by the differences between the predicted gluon PDFs at large momentum fractions $x$, which are shown in Fig.~\ref{Fig:xgluon_pdfvar} for $Q^2 = {m_t^{\mathrm{pole}}}^2 = (172\,\mathrm{GeV})^2$. The minimal $x_{min} = \mathrm{min}(x_1,x_2)$ and the maximal momentum fraction $x_{max} = \mathrm{max}(x_1,x_2)$ carried by the incoming partons have a peak at $x_{min}=0.02$ and $x_{max}=0.07$, when requiring that the event leads to a $\rho_\mathrm{s}$ value in the bulk of the distribution, explicitly in the interval $\rho_\mathrm{s} \in [0.14,0.65]$. For these values the gluon PDFs in Fig.~\ref{Fig:xgluon_pdfvar} show good agreement among each other. On the other hand, in the region of low $\rho_\mathrm{s}$, $\rho_\mathrm{s} \in [0,0.14]$, peaks are observed for $x_{\mathrm{min}}=0.15$ and $x_{\mathrm{max}}=0.25$. In fact, for this range of $x$-values a clear deviation of the predictions from different PDF sets is found, which is greater than the corresponding PDF uncertainties associated to each PDF fit.

\begin{figure}[htb]
\centerline{%
\includegraphics[width=11cm]{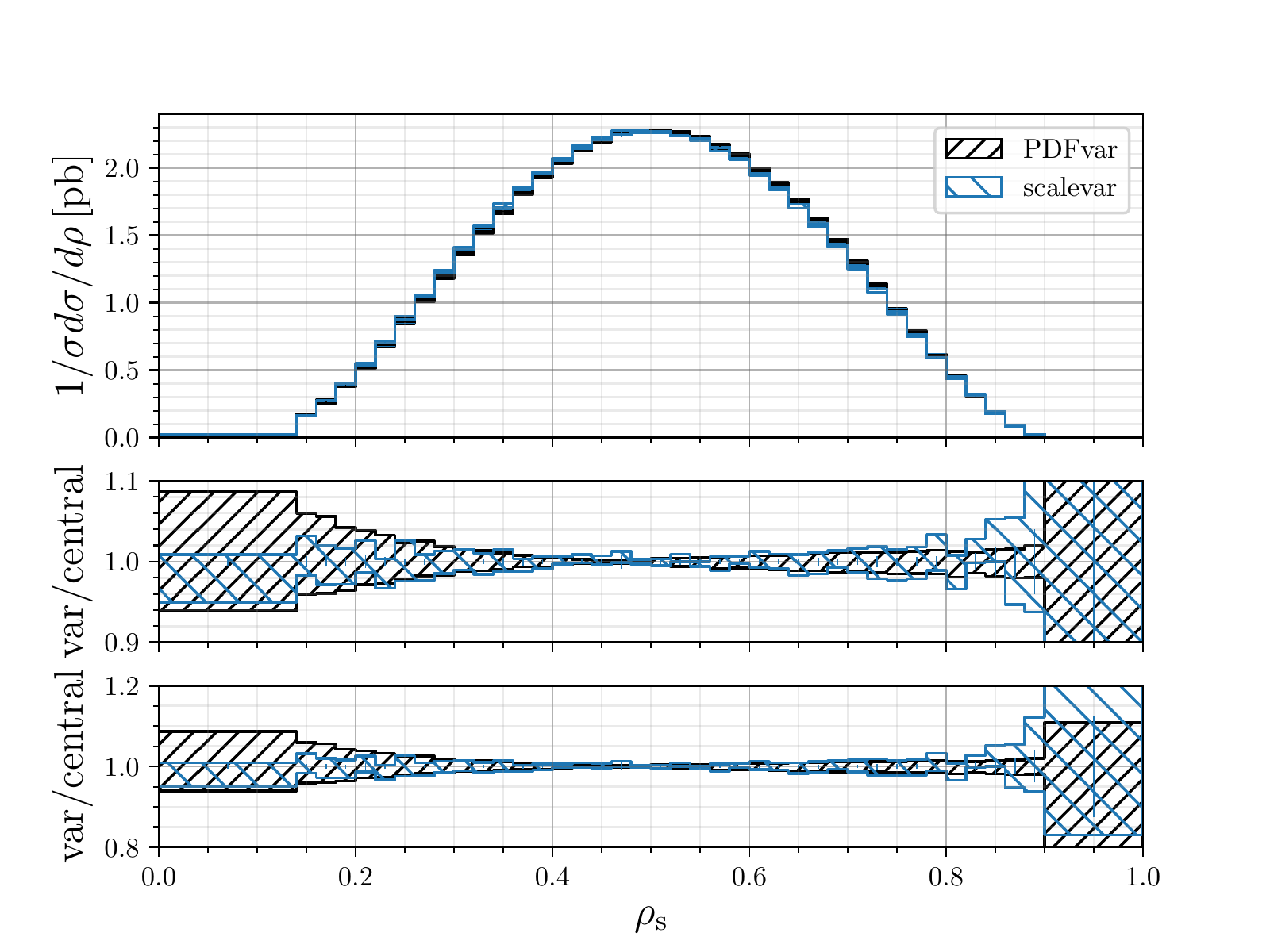}}
\caption{Normalized NLO differential cross section of the process $pp \rightarrow t\bar{t}j +X$ at $\sqrt{s}=13\,$TeV as a function of $\rho_\mathrm{s}$ obtained with the dynamical  scale $\mu_0= H_T^B/4$. The NLO seven point scale variation uncertainty band (blue) and the approximate NLO PDF uncertainty band (black) are shown, which are rescaled in the lower panel to the nominal distributions (corresponding to LO for approximate NLO PDF uncertainty and NLO for scale variation uncertainty).}
\label{Fig:rho_pdfvar_vs_scalevar}
\end{figure}

Finally, in Fig.~\ref{Fig:rho_pdfvar_vs_scalevar} the normalized $\rho_\mathrm{s}$ distribution is shown, which was obtained with the scale $\mu_0=H_T^B/4$. Both the approximate NLO PDF uncertainty (black), calculated as described above with the LO partonic cross sections and the NLO PDF set, and the NLO scale variation uncertainty bands (blue), are depicted. The PDF uncertainty in the bulk of the normalized $\rho_\mathrm{s}$ distribution is of similar size as the scale variation uncertainty for the choice of the dynamical scale $\mu_0=H_T^B/4$.

\section{Conclusions}
We investigated the theoretical uncertainties in the $t\bar{t}j$-process especially focusing on the $\rho_\mathrm{s}$ distribution, which can be used to extract the top-quark mass from experimental measurements. Thereby we studied different scale definitions and found, in agreement with \cite{Bevilacqua:2016}, that a dynamical scale is better suited to describe this process in the high-energy tails, compared to the static scale $\mu_0=m_t^{\mathrm{pole}}$. This is seen by the reduced width of the scale variation uncertainty bands in this kinematic region. Furthermore, a strong reduction of scale uncertainty in the normalized $\rho_\mathrm{s}$ distribution was observed using the dynamical instead of the static scale definition, due to the smaller shape variation induced by the scale variation. When comparing the dynamical scales, by considering the NLO and LO scale variation uncertainty bands, the scales $\mu_0=H_T^B/2$ and $\mu_0=H_T^B/4$ are found to be preferable over $\mu_0=m_{t\bar{t}j}^B/2$, as overlapping bands are found over the whole $\rho_\mathrm{s}$ range in case of the former two. Owing to the strongly reduced scale variation uncertainty in the normalized $\rho_\mathrm{s}$ distribution using $\mu_0=H_T^B/4$, this theoretical uncertainty is similar in size to the PDF variation uncertainty in the kinematical region of interest for the experimental extraction of $m_t^{\mathrm{pole}}$. Also, the influence of the $R$-parameter in the anti-$k_T$ jet clustering algorithm was investigated and was found to be negligible on the scale variation uncertainty when using the preferred dynamical scales $\mu_0=H_T^B/2$ and $\mu_0=H_T^B/4$.

\section{Acknowledgments}
We are grateful to Simone Alioli and Alessandro Gavardi for sharing with us their latest POWHEG-BOX implementation of the $t\bar{t}j$ hadroproduction process and for useful discussions, and to Adrian Irles and Peter Uwer for further discussions, suggestions, and cross-checks.

The work of M.V.G. and S.-O.M. was partially supported by
the Bundesministerium f\"ur Bildung and Forschung (contract 05H21GUCCA). The work of K.V. was partially funded by the House of Young Talents Siegen.

{\small 
\bibliographystyle{JHEP}
\bibliography{KVoss_MTTD2021}
}

\end{document}